\documentclass[twocolumn]{aa}
\usepackage{graphicx,amssymb,natbib}
\def\pl{{pl}}
\def\compps{{compps}}

\def\chiq{{$\chi^2$}}
\def\bb{{bb}}
\def\gs{{Gaussian}}
\def\nh{{$N_{\rm H}$}
\def\dbb{dbb}}

\def\B{{\em BeppoSAX }}

\begin{document}

\title{The discovery of hard X--ray emission in the persistent\\ 
flux of the Rapid Burster} 

\author{M. Falanga\inst{1}\fnmsep\thanks{ 
\email{mfalanga@cea.fr}}, R. Farinelli\inst{2},
  P. Goldoni\inst{1}, F. Frontera\inst{2,3}, 
  A. Goldwurm\inst{1}, and L. Stella\inst{4}}

 \offprints{M. Falanga}
   \authorrunning{M. Falanga et al.}
   \titlerunning{Persistent emission of the RB}

\institute{CEA Saclay, DSM/DAPNIA/Service d'Astrophysique (CNRS FRE
   2591), F-91191, Gif sur Yvette, France
\and 
Dipartimento di Fisica, Universit\`a di Ferrara, via Paradiso
  12, 44100 Ferrara, Italy
\and 
IASF, CNR, Sezione di Bologna, via Gobetti 101, 40129 Bologna, Italy 
\and 
Osservatorio Astronomico di Roma, via Frascati 33, 00040
  Monteporzio Catone, Italy} 


\abstract{
We report the first detection with {\it INTEGRAL} of persistent hard
X--ray emission  
(20 to 100 keV) from the Rapid Burster (MXB 1730--335), and describe its full
spectrum from 3 to 100 keV.
The source was detected on February/March 2003 during one of its
recurrent outbursts. 
The source was clearly detected with a high signal to noise ratio
during the single pointings and is well distinguished from the neighboring source
GX 354--0. The 3 -- 100 keV X--ray spectrum
of the persistent emission is well described by a two-component model
consisting of a blackbody plus a power-law with photon index
$\sim$ 2.4. The estimated luminosity
was $\sim$ $8.5 \times 10^{36}$ erg $s^{-1}$ in the 3 -- 20 keV energy
band and  $\sim$ $1.3 \times 10^{36}$ erg $s^{-1}$ in the 20 -- 100
keV energy range, for a distance of 8 kpc. 

 }

   \maketitle

\section{Introduction}
The Rapid Burster (MXB 1730--335, hereafter RB) is a low
mass X--ray binary system (LMXB)  located in the globular cluster
Liller 1 at a distance of $\sim$ 8 kpc \cite{ortolani96} in the Galactic plane (Lewin et
al. 1995, for a review). Unique in the Galaxy, the RB is the only LMXB known to
produce both type I and type II X--ray bursts. Type I bursts are
characteristic of LMXBs hosting a low magnetic field neutron star (NS)
and are known to be the result of explosive
thermonuclear burning of accreted material on the surface of the
NS. Type II bursts likely arise from the release of gravitational potential
energy during sudden brief periods of higher accretion rate believed
to be caused by a recurring accretion instability (e.g. Lewin et al. 1995). The RB is a
recurring transient with outbursts which last for a few weeks followed
by quiescent intervals. These intervals lasted for $\sim$ 200 days until 1999, but have 
shrinked to $\sim 100$ days since 2000 (Masetti 2002).

One of the major problems in studying the persistent emission from this source 
is its proximity (0.5$^\circ$ apart) to the bright and variable 
LMXB 4U 1728--34 (GX354--0). As a consequence, it has never been
possible to determine the persistent emission from the RB in the hard X--ray energy range, due
to the low angular resolution of the collimator detectors used so far.
Among the most recent results on the RB, we wish to mention those obtained
with the Proportional Counter Array (PCA) aboard the {\it Rossi X--ray Timing Explorer} 
({\em RXTE}) satellite \cite{guerriero99}, and those obtained with the \B\
satellite \cite{masetti00}. Guerriero et al (1999) monitored four
outbursts of the source occurred between 1996 and 1998, detecting type
I bursts, type II bursts,  
and the source persistent emission up to 20 keV. The persistent
emission level was estimated 
offsetting the satellite pointing by 0.5$^\circ$ away from GX354--0. 
They found that the 2.5 -- 20 keV persistent emission spectrum was
well described  by a multi-component model consisting of two
blackbodies ({\sc \bb}), plus a power-law ({\sc \pl}),
noticing that up to 10 keV the {\sc pl} is not needed.
Masetti et al. (2000) mainly studied in the broad 0.1 -- 200 keV energy
band the evolution of the spectral properties of the bursting emission. 
The persistent emission spectrum was investigated in the range
from 1 to 10 keV, also finding that  a 2{\sc \bb} model was suitable
to describe 
the data. At energies higher than 10 keV, a residual contamination 
from GX 354--0 prevented an unbiased estimate the persistent emission flux
level and spectrum from being performed with the Phoswich Detection System 
(PDS). Mahasena et al. (2003) observed type II bursts from th RB with {\em ASCA} in 1998 and 1999, and found that both the burst and persistent emission spectra, in the 1 -- 10 keV energy range, could be fitted with a two-component model consisting of a multi-color disc model and a blackbody.  

We observed the RB in hard X--rays, using both the imager IBIS \cite{Ubertini03} and 
the module 2 of the X--ray monitor JEM-X \cite{Lund03} on board the 
{\it INTEGRAL} satellite \cite{winkler03}.  Thanks to their high angular resolution 
it was possible to separate the RB from GX 354--0. This represents a significant
improvement over the previous measurements made with low spacial resolution
instruments. Thus we were able to clearly measure the hard X--ray
emission from the RB as distinct from that of the nearby GX 354--0.

We report here the first high energy detection of the RB during the
decay phase from outburst to quiescence. 
The observation was obtained during the February/March 2003 
outburst. The results are based on the 20 -- 100 keV data
from IBIS, and on 3 -- 20 keV data from JEM-X. In Sec. 2 we
describe the observation, in Sec. 3 we present the spectral results,
and in Sec. 4 we give the conclusion. 

\section{{\it INTEGRAL} observation and data analysis}
\label{sec:obs}

\begin{figure}[h]
   \centering
   \includegraphics[width=4.5 cm, angle=-90]{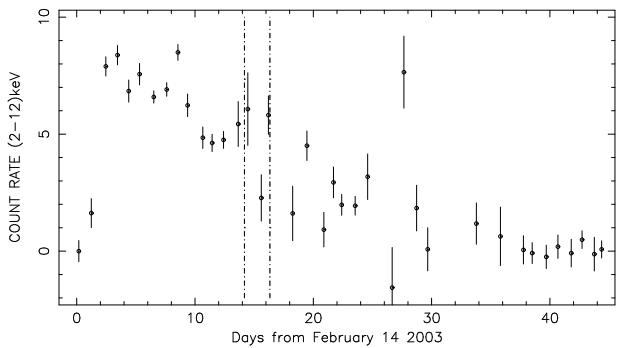}
   \caption{{\it RXTE}/ASM light curve in the 2 -- 12 keV energy band
     (data averaged over 1-day intervals) of the
     February/March outburst of the RB. The vertical dashed lines indicate the
   times of the {\em INTEGRAL} observations.}
              \label{img:lc_asm}
    \end{figure}

The present dataset was obtained during a Target of Opportunity (ToO)
observation of the Galactic center region, performed from February 28
to March 1, 2003 (satellite revolution 46), during the outburst decay phase of
the RB.  
The observation spanned just over 2 days and caught the object as the flux
level was decaying from an X--ray intensity corresponding to about 50\%  of the
outburst peak down to quiescence. 
Fig. \ref{img:lc_asm} shows the 2 -- 12 keV RB light curve obtained with the
{\it All-Sky Monitor} (ASM) on board {\em RXTE}. Also shown  in
Fig. \ref{img:lc_asm} are the time intervals in which the {\it INTEGRAL} 
observation was performed. 

The IBIS imager is a coded mask instrument with an angular resolution of 
12$'$ FWHM, a fully coded field of view  (FOV) of $9^{\circ} \times 9^{\circ}$, and
a partially coded FOV of $29^{\circ} \times 29^{\circ}$ at zero
sensitivity. The detection plane consists 
of two detection layers, ISGRI and PICsIT. The upper detector, ISGRI 
\cite{lebrun03}, has a passband 
between 15 keV and 1 MeV, but its best sensitivity is achieved
between 20 keV and 200 keV. The bottom detector, PICsIT \cite{dicocco03}, has
a passband between $\sim$ 200 keV and $\sim$ 8 MeV.
The JEM-X monitor is also a coded mask telescope with angular resolution of
3$'$ FWHM, a FOV of $4.8^{\circ}$ in diameter, and a partially coded FOV of 
$13.2^{\circ}$ in diameter. It  covers the energy range from 3 to 25 keV. 

Our observation consisted of 76 stable pointings (Science Windows,
ScW) of $\sim$ 2.2 ks exposure each, with a 5 $\times$ 5 dithering pattern. 
Due to the dithering, which implied a change of the pointing 
direction, the RB was not always 
in the JEM-X field of view. The RB position offsets, which ranged
from $1^{\circ}$ to $9^{\circ}$, were used in the data 
extraction. For IBIS, the total exposure time  was 
$\sim$ 173 ks. For JEM-X, for which the data extraction was restricted to 
pointings with an offset $\la 3^{\circ}$, the total exposure time was  $\sim$ 46 ks.
Due to the high angular resolution of IBIS and JEM-X the RB was well
separated from GX 354--0. The spectrometer (SPI) was not used since
its $2^{\circ}$ angular resolution is inadequate in this case. 

The data reduction was performed using the standard Offline Science
Analysis (OSA) version 3.0 distributed by the {\it INTEGRAL} Science
Data Center \cite{courvoisier03}. The algorithms used in the
analysis are described by  Goldwurm et al. (2003)\nocite{goldwurm03}. 
For all instruments, the newest available (OSA 3.0) response matrices 
were used. For each ScW we extracted 16 channel spectra for IBIS and
256 channel spectra for JEM-X. 
Single pointings were deconvolved and analyzed separately, and then
mosaicked.
For each ScW, source positions are determined by fitting a 2D gaussian 
function to each peak in the deconvolved image. 
Source spectra are then obtained by a simultaneous fit of all sources 
detected in the image together with the background.
The spectrum averaged on the total exposure time was obtained from those
of the ScWs by weighting them for the exposure time of
the individual pointings. 
The spectra were analyzed with the XSPEC v. 11.2.0 software package \cite{Arnaud96}.
  \begin{figure}[htb]
   \centering
   \includegraphics[width=7.5 cm]{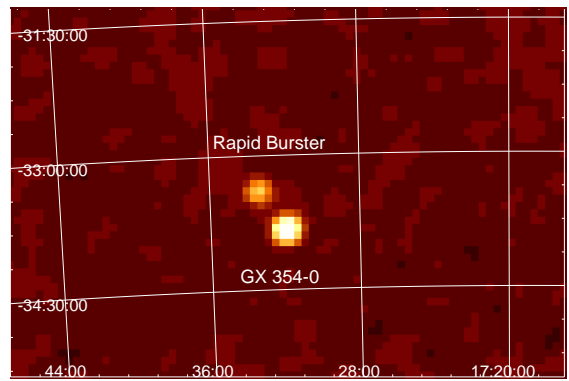}
      \caption{The 20 -- 40 keV IBIS/ISGRI mosaicked, deconvolved sky
	image of the $\sim$173 ks observation. Image size is
	$\sim6^{\circ}\times4^{\circ}$, centered at 
the RB position. The pixel size is 5$'$. The RB and GX 354--0 were
detected at a significance of $\sim$37.7 $\sigma$ and at $\sim$ 96
$\sigma$, respectively.} 
         \label{img:fig}
   \end{figure}

\section{Results}

\subsection{Image and light curve properties of the RB}

\subsubsection*{IBIS}

Figure \ref{img:fig} shows a significance map of the sources detected
in  the region 
around the RB position in the 20 -- 40 keV energy range. Two sources
are clearly detected at a 
significance level of $\sim 37.7\sigma$ for the RB and $\sim 96\sigma$ 
for GX 354--0, respectively. At higher energies, 40 -- 80 keV, the
confidence level was  $\sim 7\sigma$ for the RB and $\sim 50\sigma$ for
GX 354--0.  
With the imaging procedure, from the IBIS data the derived
celestial position of the RB is given by (J2000) $R.A. =
17^{h}33^{m}23^{s}.81,  
Dec. = -33^{\circ}23^{m}36^{s}.4$, while that of GX 354--0 is given by $R.A. =
17^{h}31^{m}57^{s}.72, Dec. = -33^{\circ}50^{m}5^{s}.6$. 
The source position offset with respect to the catalog position is
$\sim 0.34'$ for the RB, and 0.07$'$ for GX 354--0. This is within the
90\% confidence level assuming the source location error given by Gros
et al. (2003). The derived angular distance between the two sources is
$\sim 32'$.

\begin{figure}[h]
   \centering
   \includegraphics[width=6 cm, angle= -90]{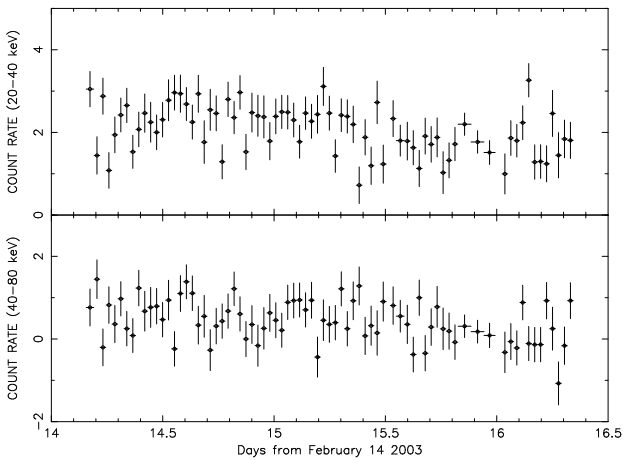}
   \caption{IBIS/ISGRI light curve of the RB in the 20 -- 40 keV ({\it top panel}) and
   40 -- 80 keV ({\it bottom panel}) energy bands. Each data point is averaged
   over one ScW, and the count
   rate was extracted directly from the source reconstructed ScW
   image. Times are expressed in days starting from 
   the outburst onset on February 14 2003.}
              \label{img:lc_ibis}
    \end{figure}

The light curve of the RB integrated on each pointing in the same energy bands
as above is shown in Fig. \ref{img:lc_ibis}.
The mean count rate in the 20 -- 40 keV and 40 -- 80 keV energy bands is 
$\sim$ 2.3 cts/s and $\sim$ 0.5 cts/s, respectively.  Above 100 keV the RB was 
not  detected at a statistically significant level either in single
exposure or in the total exposure time.
The IBIS spectrum and background are fitted simultaneously and extracted for 
each pointing and energy band. 
We excluded all channels below 20 keV and above 100 keV. 

\begin{figure}[h]
   \centering
   \includegraphics[width=6 cm, angle=-90]{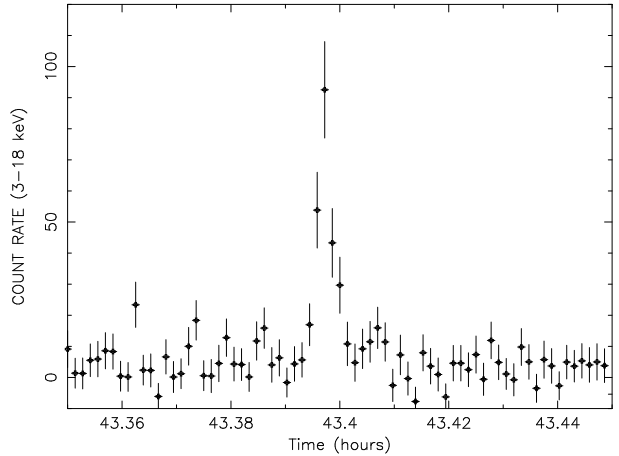}
   \caption{JEM-X 3-18 keV count rate light curve of the strongest burst. Time
 is expressed in hours from the start of the observation. The bin time
 is 5 s.}
    \label{burst}
\end{figure}

\subsubsection*{JEM-X}
\label{s:jem}

At offset angles $\la 3^{\circ}$ the RB was detected in each ScW
at significance levels of $\sim$15 -- 37$\sigma$, depending
on the angular distance from the spacecraft pointing direction. 
Thanks of the hight angular resolution (3') of JEM-X, the RB is not contaminated from GX 354--0.
The derived light curve, in addition to the persistent emission, shows 11 bursts of duration ranging from 10 s to 60 s.
The 3 -- 18 keV count rates at the burst peak range from $\sim$  29 cts/s
to 89 cts/s (see the strongest burst in Fig. \ref{burst}).

The statistics of the data however did not allow us to use the hardness
ratio to determine whether the bursts are of type I or II.
To get a reasonable signal to noise ratio, the burst light curves were
binned at 5 s, and this prevented us from looking at possible
"ringing" effects in the burst timing profile, which are characteristic of type II bursts 
(see e.g. Masetti et al. 2000). 
Because of multiple data gaps in the JEM-X light curve of the RB, we could not
establish whether the expected linear relation between the X--ray energy in
a burst and the time interval to the following burst ('E - $\Delta$t'
relation, Lewin et al. 1976\nocite{lewin76}) is satisfied by our
burst data. For the same reason, we cannot easily classify the detected 
bursts according to the results by Guerriero et al. (1999), which 
predict type I bursts at particular epochs from the outburst onset. 
On the basis of the luminosity of the persistent emission ($<2 \times 10^{37}$ erg $s^{-1}$) in correspondence of the observed bursts,
according to Guerriero et al. (1999), type II bursts are favored. However,
the estimated the 3 -- 18 keV $\alpha$  ratio between the persistent 
and bursting fluence emission (46$\pm$2), which is well above the minimum 
value found in the case of type I X--ray bursts \cite{lewin95}, would point to
a type I burst origin of the observed events.

\subsection{Persistent emission and X--ray burst spectra}
\label{sec:spec}

\begin{figure}[h]
 \centering
   \includegraphics[width=5.5 cm, angle=-90]{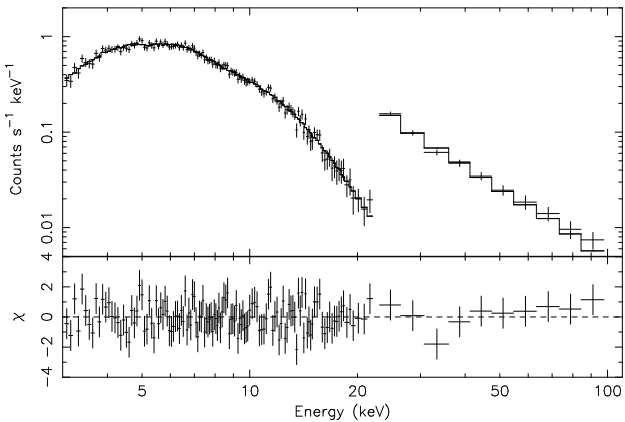}
      \caption{The 3 -- 100 keV count rate spectrum of the RB with the
        best fit model {\sc
        \bb \,+ \gs \, +  \pl}.  
        In the bottom panel are shown the residuals between the data and
 the model in units of sigma.} 
         \label{fig_spec1}
   \end{figure}

\begin{figure}[h]
 \centering
   \includegraphics[width=8.9 cm, height=5 cm]{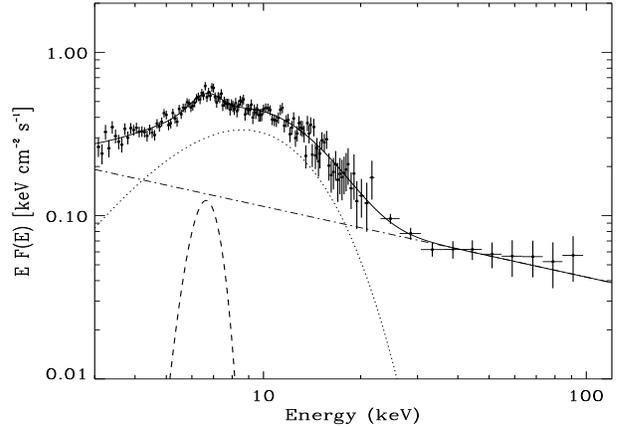}
   \vspace{0.3 cm}
      \caption{Unabsorbed 3 -- 100 keV EF(E) spectrum of the RB
	along with the  
best fit model {\sc \bb \,+ \gs \, +  \pl}. Different line styles show the 
single model components. {\it Dotted}: {\sc \bb}. {\it Dotted-dashed}:
	{\sc \pl}. 
{\it Long-dashed}: {\sc \gs}.}
         \label{fig_spec2}
   \end{figure}

The mean 3 -- 100 keV spectrum of the Rapid Burster persistent emission
was obtained by using all the useful JEM-X and
IBIS data except those corresponding to the 11 time intervals in which
the X--ray bursts were observed. 
A constant factor was also included in the fit to take
into account both the uncertainty in the instruments cross-calibration
and the fact that IBIS and JEM-X have different exposure
times: the last point in fact can introduce systematic
effects in the fitting result, expecially if the source
is characterized by a  bursting behaviour at the time
of the observations.
We fixed this constant equal to 1 for JEM-X, allowing it to vary for IBIS.
We found that the spectrum is well described
by a photo-electrically absorbed multi-component model consisting of a 
{\sc \bb} plus a {\sc \pl} plus a {\sc \gs} emission line centered 
at 6.5 keV. We were not able to constrain the interstellar column density
\nh, so \nh was frozen at 1.6 $\times$ $10^{22}$ 
cm$^{-2}$ in our fit, the Galactic value reported in the radio maps of 
Dickey \& Lockman (1990) along the source direction. This value is 
consistent with that measured by Masetti et al. (2000).
The significance of the emission line is very high, as also testified
by the \chiq/(degrees of freedom, dof) = 178/138 in the case the line 
is not included in the fit to \chiq/dof = 115/135 after its inclusion,
with a probability of chance improvement of $\sim 10^{-14}$.
The best fit parameters of the model together with the 90\% confidence
single parameter errors are reported in Table 1, while the count rate
and the $EF(E)$ spectrum
along with the single fit components and the residuals on the count
spectrum to the best fit model are shown in Fig. \ref{fig_spec1} and
Fig.\ref{fig_spec2}.

Assuming that the {\sc \pl} is due to Comptonization of soft photons off high 
energy electrons, although we do not find any evidence for a
high-energy cut-off at 100 keV, we attempted to put a lower limit on
the temperature of the electron responsible for hard X--ray emission.
For that, we substituted the simple {\sc \pl}, with the Comptonization
model {\sc \compps} of Poutanen \& Svensson (1996), assuming
a spherical geometry and a multicolour disk blackbody (Mitsuda et al.
1984) seed photon spectrum. 
This model also provides a very good description of the data
(\chiq/dof = 114/133) and gives 
a 2$\sigma$ lower limit of  90 keV for the electron temperature, an opacity
$\tau \sim 0.8$ and a inner edge disk temperature of $\sim 1.3$ keV.
Using the JEM-X data, we have also analyzed the X--ray burst spectrum.  
Because of the low statistics available for 
the single burst spectra, we summed them together deriving a mean
spectrum. 
We found that a photo-electrically absorbed {\sc \bb} describes the
data well (\chiq/dof = 65/57), with a temperature 
$kT_{bb} = 2.1 \pm 0.1$, which is consistent with that estimated for 
the persistent emission (see Table \ref{spetab}), 
and a corresponding  {\sc \bb}  radius of $R_{bb} = 3.8^{+0.3}_{-0.3}$ km.
From the {\sc \bb} model, we estimated an average burst luminosity 
of $\sim 4 \times 10^{37}$ erg s$^{-1}$. 
However, the strongest burst  reached a bolometric luminosity of
$\sim 2 \times 10^{38}$ erg s$^{-1}$.

To compare our spectral results with those obtained by Masetti
et al. (2000), we also considered the X--ray burst spectrum with
the persistent X--ray emission subtracted. 
In this case, no significant change was observed in the {\sc \bb} temperature,
and {\sc \bb} radius, while the mean burst bolometric luminosity is a little
bit lower ($\sim 3 \times 10^{37}$ erg s$^{-1}$).

\begin{table}[htb]
\begin{center}
\caption{\label{spetab} Spectral fit with blackbody, Gaussian $\&$ power
  law components.} 
\begin{tabular}{lc@{}}
\hline\hline
Parameter             & Value \\
\hline
$N_H^{a}$                    & $1.6\times 10^{22}$  cm$^{-2}$ \\
Blackbody $T_{bb}$       & $2.2^{+0.1}_{-0.1}$ (keV)\\
Blackbody $R_{bb}^{b}$   & $1.4^{+0.1}_{-0.1}$  (km)\\
Gaussian   $E_l$         & $6.5^{+0.2}_{-0.2}$ (keV)\\   
Sigma      $\sigma_l$    & $0.7^{+0.3}_{-0.3}$  (keV)\\
Equivalent Width         & $470^{+200}_{-150}$ (eV)\\
Photon index $\Gamma$         & $2.4^{+0.1}_{-0.1}$\\
$\chi^2_{\rm red}$            & 0.85 (135 d.o.f.)  \\
$L_{x}^{b}$ (3 - 100 keV)   & $9.8 \times 10^{36}$ erg s$^{-1}$ \\
$L_{\rm PL}/L_{\rm tot}$ (3 - 100 keV)    & 0.45\\
\hline\hline
a   Fixed for the fit.\\
b    Assuming a distance of 8 kpc.
\end{tabular}
\end{center}
\end{table}

\section{Discussion and Conclusion}
\label{sec:concl}

{\it INTEGRAL} has provided the first measurement of the persistent
high energy  X--ray emission from the RB without contamination from the source
GX 354--0 located at $\sim 30'$ from the RB.
We observed  X--ray emission from 3 keV up to 100 keV in the time period 
from 14 to 16 days from the outburst onset. 
The broad band persistent emission spectrum 
measured during the {\em INTEGRAL} ToO observation, is well described
by the sum of a {\sc \bb} a {\sc \gs} plus a {\sc \pl} model. The
inferred {\sc \bb} temperature is $\sim$2.2 keV, 
while the photon 
index of the high energy {\sc \pl} component is $\sim$ 2.4.
The persistent emission source luminosity is $\sim$ $10^{37}$ erg
$s^{-1}$ in the energy range from 3 -- 100 keV.  
No evidence for a cutoff in the power law component was found up 
to energies of $\sim$ 100 keV. Assuming for the hard spectral component
a thermal Comptonization model, we get a lower limit for the
electron temperature of $\sim$ 90 keV.
However we cannot exclude that the hard X--ray spectral component  
is due to non-thermal Comptonization processes occurring in 
the source (see e.g. Coppi, 1999, for a review).

An interesting feature of the RB 3 -- 100 keV spectrum 
is that, when modeled by a {\sc \bb} plus {\sc \pl}, it is very similar
to that of atoll sources in the hard state. Indeed, in some of these
sources, during their hard states, the X--ray spectrum extends 
up to several hundreds of keV without any evidence of cutoff 
(e.g., KS 1731--260 and GX 354--0, Barret \& Vedrenne 1994; Aql X--1,
Harmon et al. 1996).
This behaviour at high energies is also similar to that
of Black Hole Candidates in the so-called power--law gamma--ray
state (Grove et al. 1998).
The similarity of the observed RB  spectrum to that of 
atoll sources in hard state would be consistent with
our classification, discussed in Section \ref{s:jem}  of the 
observed X--ray bursts as  Type I bursts.
However in the atoll sources in their hard state the luminosity 
of the {\sc \bb} component is less than about 20\% with respect to their total
luminosity, while in our case the {\sc \bb} contributes to more 
than 50\% of the total luminosity.
A strong contribution of the {\sc \bb} component to the total
X--ray luminosity is instead typical of Z-sources 
(see e.g. Di Salvo et al. 2000).
This ``anomaly'' of the RB spectrum with respect to that
of the other atoll sources could however be related to the
particular, and still not well understood, physical environment
of the source (perhaps the ``propeller-effect'' caused by the magnetic field of the neutron star).  
Indeed we do not know why the RB, and only the RB, displays both
Type I and Type II X--ray bursts in continously changing times and modality.
Thus the fact that the RB cannot be fully classified as a 
``classical'' atoll source is not surprising, given the
peculiarity of the RB.
However, with {\it INTEGRAL}, a new step towards understanding
the source has been taken.
A step forward could be made by  
extending the spectral study at energies well above 100 keV,
in order to search for any cutoff or to test whether
non-thermal Comptonization processes (well observed and studied
in Black Hole Candidates) also occur in this unique source. 
This can be achieved with dedicated INTEGRAL observations of another,
possibly stronger, outburst of the Rapid Burster.

\begin{acknowledgements}

MF is grateful to E. W. Bonning for valuable suggestions and
assistance with the manuscript, and acknowledges financial support
from the France Spatial Agency (CNES).   
\end{acknowledgements}

\end{document}